\documentstyle {article}
\input epsf
\begin{document}

\begin{flushright}
KEK CP 047  \\
KEK Preprint 96-17  \\
CPTH S451. 0596  \\
\end{flushright}
\renewcommand{\baselinestretch} {1.5}
\large
\begin{center}
{\Large {\bf
Calculation of $\gamma\pi \rightarrow \pi\pi$ }}

\vskip 0.3in

 Tran N. Truong$ ^{(a,b,c)}$
\vskip 0.2in
$^{(a)}${\it National Laboratory for High Energy Physics (KEK)
\vskip 0.03in
Oho 1-1 Tsukuba, Ibaraki 305, Japan
\vskip 0.2in
$^{(b)}$Centre de Physique Theorique de l'Ecole Polytechnique
\vskip 0.03in
91128 Palaiseau, France *
\vskip 0.2in
$^{(c)}$ Department of Physics, Hue University
\vskip 0.03in 
Hue, Vietnam}
\end{center}
\vskip 0.5in

\begin{center}
{\large\bf Abstract}
\end{center}

The problem of $\gamma  \pi \to \pi \pi$ is studied using the axial anomaly,
elastic unitarity, analyticity and crossing symmetry.
  The  solution of the integral equation for this amplitude
 is given by an iteration procedure.The final solution 
disagrees with vector meson dominance models.
\vskip 0.7in
*{\it Permanent address. 

 email address: truong@pth.polytechnique.fr}

\eject

\noindent

One of the fundamental calculation in particle theory is the $\pi^0 \to \gamma
\gamma$ decay rate. It is a combination of Partial
 Conserved Axial Current (PCAC) and of 
the short distance behavior of Quantum Chromodynamics (QCD) \cite{Adler1}.
 The result of this calculation  supports  the concept of the color in QCD.
This calculation is valid in the chiral world where 
the $\pi^0$ is massless. Some correction has to be made, in principle,
 to take into account of the physical pion mass. It turns out that this axial 
 anomaly formula is in very good agreement with the pion life time data,
implying that the correction to the physical pion mass is very small. This 
result is expected because the pion mass is small.

Another axial anomaly result is the process $\gamma \pi \to \pi\pi$ or
 its analytical continuation $\gamma \to 3\pi$ \cite{Adler2}.
 This last process, even at moderate energy has a complex singularity
and requires extra work and is postponed to a forthcoming study.

The calculation of the process  $\gamma \pi \to \pi\pi$  is in itself interesting
 because not only it will be checked
 by future experiments which are being
proposed at various accelerator facilities, but it will also
 be important for the calculation  $\pi^0 \to \gamma \gamma^*$ \cite{Truong1}.

Omitting the kinematical factor,
the $\gamma \to 3\pi$ and $\gamma \pi \to \pi\pi$ amplitudes are given as:
\begin{equation}
F_{3\pi}(0) = \frac{1}{4\pi^2 f_\pi^3} = \lambda \label{eq:anomaly3}
\end{equation}
where $f_\pi=93$ MeV and the zero in the argument of $F_{3\pi}$ refers to the massless pions, 
and to the amplitude  evaluated  at the symmetry point $s=t=u=0$. 

Using the KSRF relation \cite{KSFR} $g_{\rho\pi\pi} = m_\rho/(\sqrt{2} f_\pi)$, 
 it is  straightforward to calculate the process $\gamma\pi \rightarrow
 \pi\pi$ \cite{Bando,Rudaz} using the vector meson dominance model (VMD)\cite{Sharp}:

\begin{equation}
F_{3\pi}^{VMD}(s,t,u) = \frac{\lambda}{2}(\frac{m_\rho^2}{m_\rho^2-s}+
\frac{m_\rho^2}{m_\rho^2-t}+\frac{m_\rho^2}{m_\rho^2-u}) \label{eq:vmd}
\end{equation}
which in the chiral limit,$ s=t=u=0$, yields $F_{3\pi}^{VMD}(0,0,0) = \frac{3}{2} \lambda$
 and is a factor 3/2 larger than the anomaly Eq. (\ref{eq:anomaly3}). It is clear  
that the VMD model does not always work. In order to rectify this problem,
 a contact term is introduced 
to reduce the VMD contribution \cite{Bando,Terentev}. The VMD equation
 can now be written as:
\begin{equation}
F_{3\pi}^{CVMD} (s,t,u) = \frac{\lambda}{2}(\frac{m_\rho^2-m_\pi^2}{m_\rho^2-s}+
\frac{m_\rho^2-m_\pi^2}{m_\rho^2-t}+\frac{m_\rho^2-m_\pi^2}{m_\rho^2-u}- 1) \label{eq:vmdC}
\end{equation}
and is seen to satisfy the anomaly Eq. (\ref{eq:anomaly3}) in the chiral limit.
 In  Eq. (\ref{eq:vmdC}) the smoothness assumption was made by setting  
$F_{3\pi}(s=t=u=m_\pi^2)= \lambda$. This assumption will be assumed to be correct in 
the remaining of this letter.

Yet another solution was discussed \cite{Rudaz}:
\begin{equation}
F_{3\pi}^{MVMD} (s,t,u) = \frac{\lambda}{3}(\frac{m_\rho^2-m_\pi^2}{m_\rho^2-s}+
\frac{m_\rho^2-m_\pi^2}{m_\rho^2-t}+\frac{m_\rho^2-m_\pi^2}{m_\rho^2-u} ) \label{eq:vmdM}
\end{equation}
and is seen to satisfy the anomaly Eq. (\ref{eq:anomaly3}) but
 differs from the Vector Meson Dominance result by a factor of 3/2. 

In this article a different result on the $F_{3\pi}$ amplitude 
 based on the solution of an integral 
equation  is given. The strong P-wave $\pi\pi$ scattering 
phase shifts is
supposed to be known as given by the experimental data up to 1 GeV or higher, i.e. the 
existence of the $\rho$ resonance at $0.77$ GeV with a width of $0.151$ GeV.
We shall not make an assumption on the Vector Meson Dominance, but try to find a 
solution which is consistent with the constraint of the elastic unitarity, crossing symmetry 
and also of the low energy theorem Eq. (\ref{eq:anomaly3}).
 Our solution does not agree either with
 Eq. (\ref{eq:vmdC}) 
or with Eq. (\ref{eq:vmdM}).

The process  $\gamma \pi \to \pi\pi$ is a
completely symmetric reaction, that is the same amplitude describes not only 
the reaction
 $\gamma \pi^0 \to \pi^+\pi^-$, but also two other
amplitudes involving the permutations of the pions. One assumes that the
 scattering amplitude can be represented by a single spectral function dispersion
relation:

\begin{equation}
F(s,t,u)= \lambda + [(s-m_\pi^2)\int_{4m_\pi^2}^\infty
\frac{\sigma(z) dz}{(z-m_\pi^2)(z-s-i\epsilon)} ] + [s \leftrightarrow t ]+ [s \leftrightarrow u] 
\label{eq:fstu} 
\end{equation}

The spectral function can be evaluated, using the unitarity relation: 
$\pi\sigma(s,t,u)= \langle\pi^0\mid j_3(0)\mid n \rangle \langle n\mid \pi^+\pi^-\rangle$
where the $j_3$ is the isovector electromagnetic current operator and $\mid n>$
is a complete set of the physical states; the summation over $\mid n>$ is understood.
 Because of the G-parity, only the 
even number of pions intermediate states contribute. 
As one is interested in the low energy region (but including possible
resonant effect) of the  $\gamma \pi^0 \to \pi^+\pi^-$ amplitudes where
 the inelastic effects are small, the summation over the intermediate states 
is truncated to the two pion (P-wave) states.
The amplitudes
 $\langle\pi^0\mid j_3(0)\mid \pi^+\pi^-\rangle$ and $\langle\pi^+\pi^-\mid \pi^+\pi^-\rangle$ are expanded 
into partial waves, and keeping only the lowest  partial wave contributions 
(physically the P-state), one has:
 \begin{equation}
\pi\sigma(s) = F(s)e^{-i\delta(s)} sin\delta(s)  \label{eq:unitarity3}
\end{equation}
where $F(s)$ denotes the S-wave projection of $F(s,t,u)$ (and physically corresponds
 to the P-wave when combining with the kinematical factor); 
    $\delta$ is the experimental P-wave $\pi\pi$ phase shifts
which can be obtained from the 
experimental data  which shows that the phase shifts must pass 
through $90^{\circ}$ at the $\rho$ mass and its
 width is $151 MeV$.

Substituting Eq. (\ref{eq:unitarity3}) in the partial wave projection
of Eq. (\ref{eq:fstu}) and keeping only the lowest partial wave, one has:

\begin{eqnarray}
F(s)&=& \lambda + {(s-m_\pi^2)\over\pi}\int_{4m_\pi^2}^\infty
\frac{F(z)e^{-i\delta(z))}\sin\delta(z)}{(z-m_\pi^2)(z-s-i\epsilon)}dz  \nonumber \\
&& +{2\over\pi}\int_{4m_\pi^2}^\infty
F(z)e^{-i\delta(z)}\sin\delta(z)
\{{1\over(s-4m_\pi^2)}\ln[1+{(s-4m_\pi^2)\over z}]-{1\over(z-m_\pi^2)}\}dz  \nonumber \\
\label{eq:int2}
\end{eqnarray}
The unitarity of the S-matrix requires that $F(s)$ has the phase $\delta$ \cite{Watson}
 and this should be
exhibited in the solution for $F(s)$.
Because of the neglect of the contribution of higher partial waves, $F(s)$ is not
 exactly equal to $\lambda$ at $s=m_\pi^2$.
Eq. (\ref{eq:int2}) is a complicated integral equation; it is similar to, but
more complicated than 
 the Muskelishvili-Omnes (MO) type \cite{Omnes}, because the t and u channel contributions
 are also expressed in terms of the unknown function $F(s)$. It should be noticed that 
the first term has a cut from $4m_\pi^2$ to $\infty$ and the second one has a cut from
 $0$ to $-\infty$. This remark enables one to solve the integral equation by an iteration
scheme as given below which converges very fast.

The iterative and final solutions can be expressed
 in terms of the function $D(s,0)$, normalized to unity at $s=0$ and defined in terms of the phase
 shift $\delta$ as:
\begin{equation}
{1\over D(s,0)} = \exp[ {s\over \pi} \int_{4m\pi^2}^\infty 
{\delta(z)dz\over {z(z-s-i\epsilon)}}]
\label{eq:omnes}
\end{equation}
Other functions $D$ normalized to unity at $s=s_0$ can be expressed in terms of
 the function $D(s,0)$ by the simple relation $D(s,s_0)=D(s,0)/D(s_0,0)$.

As remarked above, the Integral Equation (\ref{eq:int2}) has both right and left cuts.
This allows us to define an iteration procedure defined as follows:
\begin{equation}
F^{(i)}(s) = \frac{\lambda}{3} + T_B^{(i-1)}(s) +  \frac{s-m_\pi^2}{\pi}\int_{4m_\pi^2}^\infty
\frac{F^{(i)}(z)e^{-i\delta(z)}\sin\delta(z)}{(z-m_\pi^2)(z-s-i\epsilon)}dz  \label{eq:inti}
\end{equation}
where $F^{i}$ is the value of the function $F(s)$ calculated at the $i^{th}$ step
 in the iteration procedure; the Born term $ T_B^{i-1}(s)$ calculated at the $i^{th}-1$ 
step is defined as
\begin{equation}
T_B^{(i-1)}(s)=\frac{2\lambda}{3}+{2\over\pi}\int_{4m_\pi^2}^\infty
F^{(i-1)}(z)e^{-i\delta(z)}\sin\delta(z)
\{{1\over(s-4m_\pi^2)}\ln[1+{(s-4m_\pi^2)\over z}]-{1\over(z-m_\pi^2)}\}dz \label{eq:tbi-1}
\end{equation}
where $i\geq 1$. The Born term is real for $s\geq 0$.

The solution of the integral equation Eq. (\ref{eq:inti}) is of the MO type \cite{Omnes}:
\begin{equation}
F^{(i)}(s)= \frac{\lambda}{3D(s,m_\pi^2)} + T_B^{(i-1)}(s) + \frac{1}{D(s,m_\pi^2)}\frac{s-m_\pi^2}{\pi}
\int_{4m_\pi^2}^\infty \frac{D(z,m_\pi^2 e^{i\delta(z)}\sin\delta(z) T_B^{(i-1)}(z) dz}
{(z-m_\pi^2)(z-s-i\epsilon)} \label{eq:sol1}
\end{equation}
where it is assumed that the well-known polynomial ambiguity inherited in the 
 MO integral equation is absent \cite{Omnes}. (The polynomial ambiguity
 could represent some incalculable inelastic effect occurring above the 
inelastic threshold). It is now straightforward to take out
 the imaginary part of the integral in Eq. (\ref{eq:sol1}) and using $e^{i\delta}\sin\delta=
\rho(s)N(s)/D(s,0)$ with $ImD(s,0)=-\rho(s)N(s)$
 to combine with the $T_B^{(i-1)}$ Born term
 to show that $F^{(i)}(s)$ has indeed the
 phase $\delta$ as required by the the unitarity of the S-matrix.

 One  arbitrarily defines the convergence of the iteration scheme at the ith iteration step 
when $\mid F^{(i)}\mid/\mid F^{(i-1)}\mid$ differs from 1 by less than 1\% or so for the
energy range
 from the 2 $\pi$ threshold to $1 GeV$. (Alternatively one can also consider the ratio 
$\mid T_B^{(i)}\mid/\mid T_B^{(i-1)}\mid$). 
 Then
combining the
$T_B^{(i-1)}$ Born term in Eq. (\ref{eq:sol1})
 with higher uncorrected 
partial waves (for rescattering) from the $t$ and $u$ channels, one arrives at the final solution:
\begin{equation}
F^{(i)}(s,t,u)= \frac{\lambda}{3}[ \{\frac{1}{D(s,m_\pi^2)}(1+3I^{(i-1)}(s))\} + \{(s\leftrightarrow t)\}
 +\{(s\leftrightarrow u)\} ]
 \label{eq:final}
\end{equation}
where the function $I^{(i-1)}$ is the rescattering correction:
\begin{equation}
I^{(i-1)}(s) = \frac{s-m_\pi^2}{\pi} \int_{4m_\pi^2}^\infty 
 \frac{D(z,m_\pi^2) e^{i\delta(z)}sin\delta(z) T_B^{i-1}(z) dz}
{(z-m_\pi^2)(z-s-i\epsilon)} \label{eq:I}
\end{equation}
By projecting out the lowest partial wave from Eq. (\ref{eq:final}) it can be shown that 
$F^{(i)}(s)$ has the phase $\delta$. 

In order to carry out the iteration procedure to find the solution of the integral equation
one has to parameterize the function $D(s,0)$ in terms of the experimental P-wave phase shift.
 Noticing that the phase of $D(s,0)$ is -$\delta$), one has 
\cite{Brown,Gounaris}.
\begin{equation}
        {1\over D(s,0)} = \frac{1} {1 -s/s_{R} - {1\over 96\pi^2f_\pi^2}\{(s-4m_\pi^2)
 H_{\pi\pi}({s}) + {2s/3}\}} \label{eq:vu}
\end{equation}
where $ H_{\pi\pi}(s)=(2-2\sqrt{s-4m_\pi^2\over s}\ln{{\sqrt{s}+\sqrt{s-4m_\pi^2}\over
2m_\pi}})+i\pi\sqrt{s-4m_\pi^2\over s}$
for $ s>4m_\pi^2$; for other values of s, $H_{\pi\pi}(s)$ can be obtained by analytic
 continuation. 
The partial wave amplitude is written as $e^{i\delta}\sin\delta/\rho(s)=N(s)/D(s,0)$,
 where $N(s)=(s-4m_\pi^2)/(96\pi f_\pi^2)$.

 The $\rho$ mass is defined as the vanishing of the real part of the 
denominator and is equal to $s_R$ in the narrow width approximation. 
Using $\sqrt s_\rho$=$0.770 GeV$ in Eq. (\ref{eq:vu}), we have $\Gamma_\rho=155 MeV$
which is very close to the experimental value of $151.5\pm 1.5 MeV$. 

To make the iteration scheme coherent, the phase theorem for $F^{(i)}(s)$ has to be
satisfied  at every step of the iteration. 
One first guesses a solution for $F^{(0)}$ which must satisfy this theorem; the 
following $F^{(i)}(s)$ satisfies the theorem by construction. One first sets: 
$F^{(0)}(s)=(\lambda/3)D^{-1}(s,m_\pi^2)$; we
 then put this solution 
 into Eq. (\ref{eq:tbi-1}) to evaluate the 0th order Born term:
\begin{equation}
T_{B}^0(s) = \frac{2 \lambda}{3} (\frac{s_\rho-m_\pi^2}{s-4m_\pi^2}) 
 \ln \mid1+\frac{s-4m_\pi^2}{s_\rho}\mid \label{eq:Born0}
\end{equation}

To arrive at this relation one takes a $\delta$-function
 approximation for the product \newline 
 $e^{-i\delta}\sin\delta(s)/D(s,0)$ which is sufficiently accurate at this point.
 Eq. (\ref{eq:Born0}) can also be obtained from projecting out
 the lowest partial wave from the t and u channels contribution of Eq. (\ref{eq:vmdM}).

Using this Born term, $F^{(1)}$ is calculated, using Eq. (\ref{eq:sol1}) with $i=1$.

One next calculates the next Born $T_B^1(s)$ term by using
 the solution $F^{(1)}(s)$ and then proceeds to calculate $F^{(2)}(s)$ etc.
It is found that $\mid F^{(1)}(s)\mid$ differs by less than 2.5\% from $\mid F^{(2)}(s)\mid$ 
and  $\mid F^{(2)}(s)\mid$ differs by less than 0.4\% from  $\mid F^{(3)}(s)\mid$  in the
energy  range of $2m_\pi$ to $1 GeV$ which shows indeed the
 very fast convergence of our iteration scheme.

If we were happy with an accuracy of about 1.5\% in the amplitude, then the final solution
given by Eq. (\ref{eq:final}) could  be simply expressed in terms of $I^{(0)}(s)$.
 The exact expression for $I^{(0)}(s)$ is a complicated linear combination of the 
Spence functions, but an approximate expression can simply be given as follows:
\begin{equation}
I^{(0)}(s) \simeq 1.13\frac{s_\rho}{144\pi^2 f_\pi^2}\{Sp(1+\frac{s-4 m_\pi^2}{s_\rho})
-Sp(1-\frac{3 m_\pi^2}{s_\rho})\} \label{eq:xy}
\end{equation}
where $Sp(s)$ is the Spence function. With this approximation, the expression for 
$F^{(1)}(s,t,u)$ is accurate to within 2\% compared to the final solution constructed from
$F^{(3)}(s)$ in the energy range of 0.1 to 1 GeV.
 
In order to compare these results with the previous pole models, one has  
to make a finite width correction for Eqs. (\ref{eq:vmdC},\ref{eq:vmdM}). This can be done by
using the self-energy correction  for the $\rho$ propagator using the $\rho\pi\pi$ coupling
given by the KSRF relation. One then has to make the substitution
$(s_\rho-m_\pi^2)/(s_\rho-s)$ by 
$1/D(s,m_\pi^2)$ and similarly for expressions with $t$ and $u$ variables. The lowest S-
wave projection yields:
\begin{equation}
F_{3\pi}^{CVMD}(s)=\frac{\lambda}{2}[\frac{1}{D(s,m_\pi^2)}+T_B^0(s)-1 ]
 \label{eq:a1}
\end{equation}
and
\begin{equation}
F_{3\pi}^{MVMD}(s)=\frac{\lambda}{3}[\frac{1}{D(s,m_\pi^2)}+T_B^0(s) ] 
\label{eq:a2}
\end{equation}
where it is sufficiently accurate to use the pole terms for the
 contribution of the $t$ and $u$
channels. The phase theorem is violated in both Eqs. (\ref{eq:a1},\ref{eq:a2}).

Fig. 1 shows that the final solution lies between the two VMD models, the
 (CMVD) given by  
Eq. (\ref{eq:a1}), and the (MVMD) model given by Eq. (\ref{eq:a2}).
 It
is useful to subject our result with future experimental test in the resonant region where
 the difference with other models are largest. 

Fig. 2 shows a similar plot but with only a low energy  scale $s<0.3 GeV^2$.
 The present experimental data  at low energy \cite{Antipov} confirms roughly the anomaly but is not
 sufficiently accurate to distinguish various models. With  more accurate data, below 
$s<0.3 GeV^2$, it will be  experimentally difficult to distinguish our result with that of
CVMD but it might be possible to distinguish our result with that of MVMD. For 
a more complete review of the experimental situation as well as other theoretical models,
 two recent articles  are to be consulted \cite{Ivanov,Roberts}.

It will be best to analyze future experimental data using Eq. (\ref{eq:final}),
 i.e. in terms of the energy and scattering angle instead of its 
 S-wave projection as given by Figs. 1 and 2.

In conclusion,the process  $\gamma \pi \to \pi\pi$ is calculated using the low energy theorem,
 analyticity, elastic unitarity and crossing symmetry which are 
fundamental conditions for a theory 
 involving strong interaction. The final result shows that one effectively takes into account of 
 the (unstable) $\rho$ model in the s, t and u
 channels and their rescattering effect treated  in a self-consistent way.
 It is important to put these results to experimental tests.

Part of this work was done while the author was a visitor at the KEK National Laboratory
 and at the Hue University. The author would like to 
thank these two institutions for their hospitality and  
 in particular, Prof. Y. Shimizu and his collaborators at KEK
 and Dr Nguyen trung Dan at the Hue University.

\newpage

{\bf Figure Captions}

Fig.1:$\mid F^{(i)}(s)\mid^2/\lambda^2$ as a function of the energy squared $s (GeV^2)$.
The solid curve is the solution of the integral equation after 3 iterations; the dotted curve 
is the solution after one iteration. The long dashed curve is from the modified
VMD model (MVMD) Eq. (\ref{eq:a2}); the short dashed curve 
is from the VMD model with a contact term (CVMD) Eq. (\ref{eq:a1}). 

Fig. 2: $\mid F^{(i)}(s)\mid^2/\lambda^2$ as given by the Fig. 1 but with a smaller energy scale.

\newpage
\begin{figure}
\epsfbox{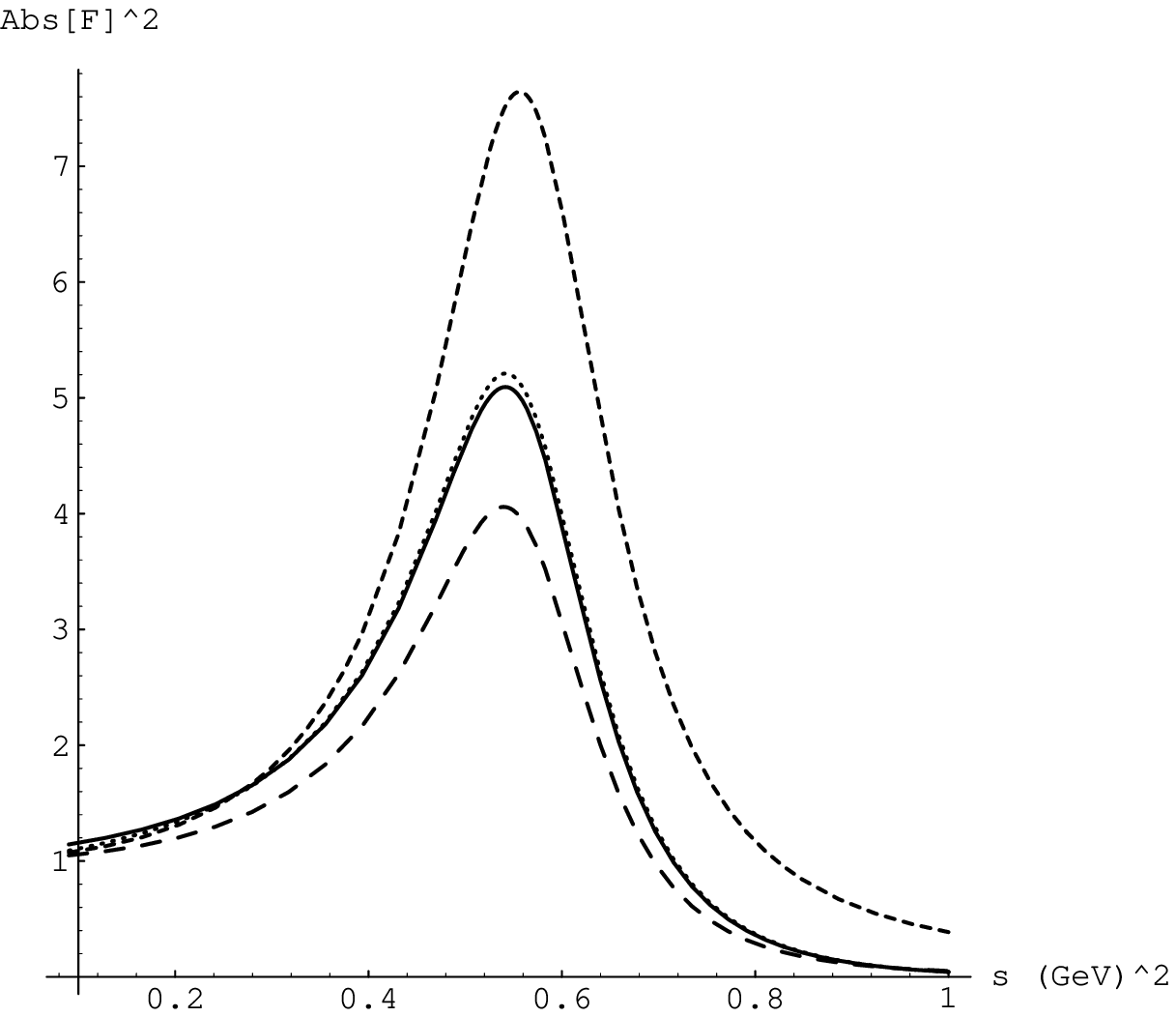}
\caption{}
\label{Fig.1}
\end{figure}
\newpage
\begin{figure}
\epsfbox{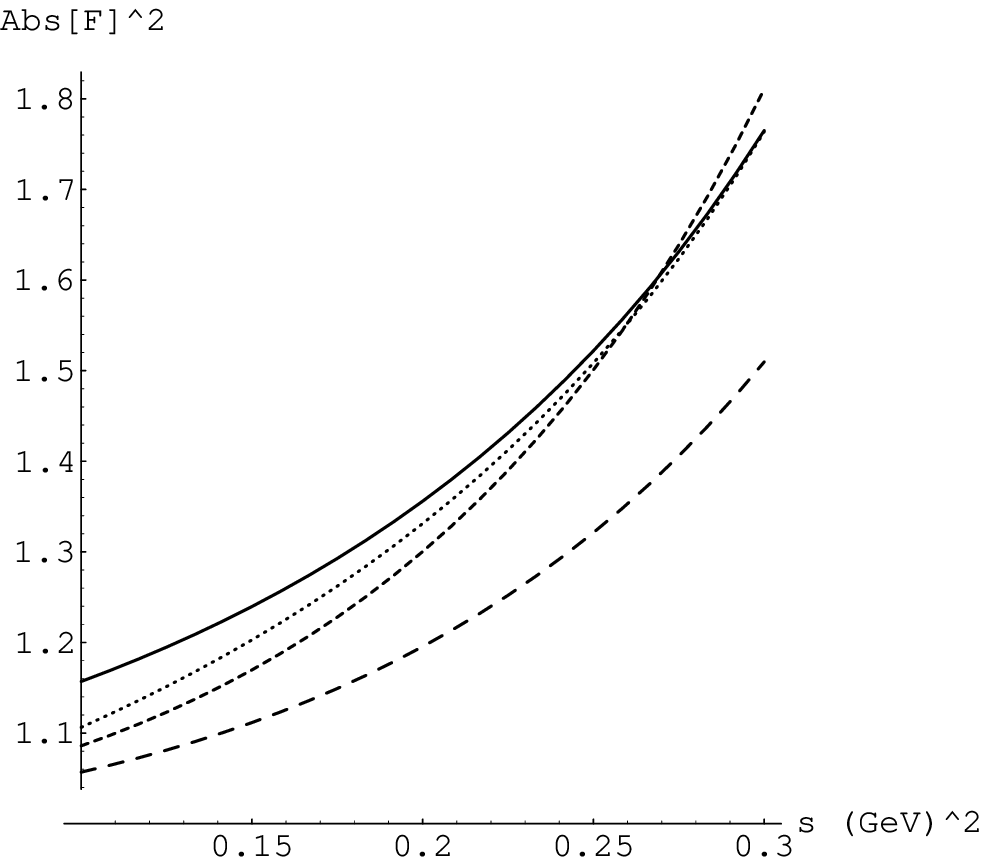}
\caption{}
\label{Fig.2}
\end{figure}

\end{document}